\documentclass[a4paper,12pt]{article}
\usepackage{ifpdf}
\ifpdf
\usepackage[pdftex]{graphicx}
\usepackage[usenames,dvipsnames]{color}
\else
\usepackage{graphicx}
\usepackage[usenames,dvips]{color}
\fi
\usepackage{amsmath}
\usepackage[colorlinks=true]{hyperref}
\usepackage{caption}
\usepackage{paper}

\numberwithin{equation}{section}

\begin{document}

\title{\bfseries Deriving the ADM 3+1 evolution equations
                 from the second variation of arc length.}
\author{%
Leo Brewin\\[10pt]%
School of Mathematical Sciences\\%
Monash University, 3800\\%
Australia}
\date{02-Nov-2009}
\reference{Preprint: arXiv:0903.5365\\[5pt]
           Journal: {\it Phys.Rev.D.} {\bf 80} 084030 (2009)}

\maketitle

\begin{abstract}
\noindent
We will show that the ADM 3+1 evolution equations, for a zero shift vector, arise naturally from the
equations for the second variation of arc-length.
\end{abstract}

\section{Introduction}
\label{sec:intro}

Discussions of the dynamics of general relativity often begin with the ADM 3+1 evolution equations
\cite{gourg:2007-01}. These equations describe the second time derivatives of the spatial metric in
terms of other quantities such as the lapse function and the extrinsic and Riemann curvatures. If by
some means we happen to have a local solution (in time) of those equations then we could imagine
computing the arc length for short geodesic segments. What then would we get for the value of the second
time-derivative of that arc-length? This question has been discussed many times elsewhere
\cite{chavel:2006-01,hicks:1965-01} but under a different name -- the second variation of
arc-length. Clearly the second variation of arc-length and the ADM 3+1 evolution equations must be
related. The purpose of this paper is to establish that relationship. The result is not un-expected --
the equations for the second variation of arc-length can be used to recover the standard ADM 3+1
evolution equations with zero shift vector.

There is value in this presentation beyond the purely pedagogical -- the results presented here provide
strong theoretical support for an approach to numerical relativity being developed by the author
\cite{brewin:2002-01,brewin:1998-02,brewin:1998-01,brewin:2009-05}. This method is known as smooth
lattice relativity and is closely related to the Regge calculus
\cite{regge:1961-01,gentle:2002-01,williams:1992-01}. Both methods use a lattice to describe the metric
but they differ most notably in the way they treat the curvatures. In the Regge calculus the metric
is piecewise flat with the result that the curvatures are distributions on the 2-dimensional subspaces
known as bones (or hinges) while on a smooth lattice we allow the metric to vary smoothly in the
neighbourhood of any vertex. This allows all the usual tools of differential geometry to be applied to
the smooth lattice. In particular we can easily compute the Riemann and extrinsic curvatures in terms of
the geodesic arc-lengths of the lattice and thus, using equations (\ref{ADMDLijb},\ref{ADMDLijc}) or
\eqref{SecondDeriv}, the second time derivatives of the leg-lengths. This makes the study of dynamics on
a smooth lattice quite simple in principle (though as with any numerical method in general relativity
the practical aspects are far from trivial). Attempts have been made to adapt the ADM 3+1 equations to
the Regge calculus \cite{friedman:1986-01,piran:1986-01} but progress has been slow. A much more
promising scheme, for the Regge calculus, is due to Sorkin \cite{sorkin:1975-01} with later development
by Barrett et al. \cite{arXiv:gr-qc/9411008v1} and Gentle and Miller \cite{arXiv:gr-qc/9706034v2}.

\section{First and second variations}
\label{sec:1st2ndvar}

Discussions on the first and second variations normally arise when asking questions about geodesics such
as: Is the geodesic that joins two points unique? Is it the shortest geodesic? How far can the geodesic be
extended before it fails to be the shortest geodesic? The mathematical theory that answers these
questions is very elegant and has previously found its way into general relativity as a tool in studying
the global properties of spacetime \cite{hawk-ellis:1973-01}. Hawking and Penrose
\cite{penrose:1965-01,hawking:1967-01} made extensive use of the first and second variations of
non-spacelike geodesics in their singularity theorems. In contrast, our interest in the first and
second variations is that they provide a natural setting in which to ask different questions of
geodesics: How can the first and second time derivatives (of the arc-length) be computed? And how are
they related to the curvature tensors? As already noted in the introduction these questions will lead to
the standard ADM 3+1 evolution equations with a zero shift vector. But first we need to introduce some
basic notation and to make clear the class of curves we will be working with.

Choose a point $i$ and a small neighbourhood of $i$ in which the spacetime is non-singular. All of the
curves we are about to construct will have a finite length and will lie totally within this neighbourhood.
Through $i$ construct a timelike curve $C_i$ with affine parameter $\eta$. From $C_i$ we can construct a
nearby curve $C_j$ by dragging $C_i$ sideways a short distance (i.e. drag $C_i$ along a short spacelike
vector field defined on $C_i$). We now have two nearby timelike curves $C_i$ and $C_j$ (the point $j$ on
$C_j$ can be easily identified -- it has the same $\eta$ value as $i$ has on $C_i$). We will assume that
the two curves $C_i$ and $C_j$ are sufficiently close that, for any given $\eta$, we can construct a
unique geodesic that joins the two curves.

\begin{figure}[t]
\includegraphics[width=\textwidth]{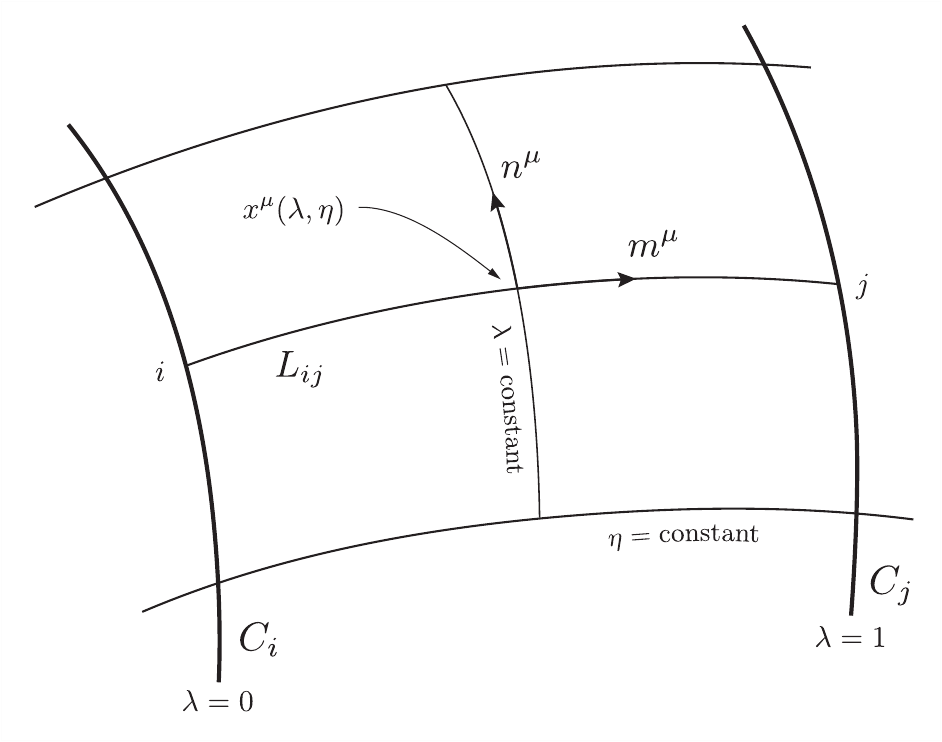}
\vskip0.5cm
\caption{\normalfont This figure displays the 2-dimensional surface $S$ constructed
from the pair of time like worldlines $C_i$ and $C_j$. The curve connecting $i$ to $j$ is a spacelike
geodesic with length $L_{ij}$. Along these geodesics $\eta=\,$constant. Note that the tangent vectors
$n^\mu$ and $m^\mu$ are unit vectors but they need not be mutually orthogonal.}
\label{LamEta}
\end{figure}

Consider now the family of geodesics generated by allowing $\eta$ to vary. This family of geodesics
(actually, segments of geodesics) will cover a 2-dimensional subspace (like a taut ribbon) which we will
denote by $S$. We will introduce coordinates on $S$ in a rather obvious way. Consider a point $P$ on
$S$. There will be exactly one space like geodesic (of $S$, by assumption) that passes through $P$. The
point $P$ will be located some fraction, $\lambda$, along the geodesic from $C_i$ to $C_j$. Thus the
points on $C_i$ will have $\lambda=0$ while those points on $C_j$ will have $\lambda=1$. We will take
the other coordinate for $P$ to be the value of $\eta$ that identifies this geodesic from all others (in
$S$). The coordinates for $P$ are then taken to be $(\lambda,\eta)$. This situation is displayed in
figure (\ref{LamEta}).

Consider now a global coordinate system, $x^\mu$, for the spacetime. Then $S$ can also be described by
functions of the form $x^\mu(\lambda,\eta)$. We now define a pair of vectors $\eta^\mu$ and
$\lambda^\mu$ by
\begin{gather}
\eta^\mu = \PxPeta{\mu}\>,\qquad
\lambda^\mu = \PxPlam{\mu}
\label{vecNMa}
\end{gather}
and a pair of unit vectors $n^\mu$ and $m^\mu$ by
\begin{gather}
n^\mu = \frac{1}{N}\eta^{\mu}\>,\qquad
m^\mu = \frac{1}{M}\lambda^{\mu}
\label{vecNMb}
\end{gather}
where $N$ and $M$ are scalar functions that ensure that the vectors are indeed unit vectors. Clearly the
vector $\eta^\mu$ is tangent to the $\lambda=\,$constant curves while $\lambda^\mu$ is tangent to the
$\eta=\,$constant curves (and both vectors will, in general, be neither unit nor orthogonal, despite
appearances in figure (\ref{LamEta})). It is rather easy to show that $M = ds/\dlam = L_{ij}$ where
$s$ is the proper distance along the geodesic and $L_{ij}$ is the length of that geodesic. Recall that
$ds/\dlam$ is constant along a geodesic while $L_{ij} = \int_0^1 (ds/\dlam)\>\dlam$ and thus $L_{ij} =
ds/\dlam$. Next, using the requirement that $m^\mu$ be a unit vector leads immediately to
$M=ds/\dlam=L_{ij}$ as claimed. Later, when we specialise to the ADM 3+1 formulation in section
\ref{sec:ADMpt1} we shall see that $N$ is the usual lapse function associated with the time coordinate
$\eta$.

We can now state clearly the equations for the arc-length and their variations.
\begin{align}
\intertext{\bf Arc-length}
\Lij &= \int_i^j\>\left( g_{\mu\nu} \PxPlam{\mu}\PxPlam{\nu} \right)^{1/2}\>\dlam
\label{ZeroDeriv}
\\
\intertext{\bf First variation}
\DLijDeta &= \left[ m_\mu \eta^\mu \right]_i^j
           = \int_i^j\>m_\mu m^\nu \eta^\mu{}_{;\nu}\>ds
\label{FirstDeriv}
\\
\intertext{\bf Second variation}
\DDLijDeta
   &= \left[ \eta^\alpha{}_{;\mu} \eta^\mu m_\alpha \right]_i^j
     - \int_i^j\> R_{\mu\alpha\nu\beta} m^\mu m^\nu \eta^\alpha \eta^\beta \>ds\notag\\
   & +\int_i^j\> \left(
          \eta_{\mu;\alpha}\eta^{\mu}{}_{;\nu}m^\alpha m^\nu
          - \left(m_\mu m^\nu \eta^\mu{}_{;\nu}\right)^2 \right) \>ds
\label{SecondDeriv}
\end{align}

For ease of reference we have included a proof of the above equations in the appendices.
See also \cite{chavel:2006-01,hicks:1965-01} for more details.

\section{The ADM evolution equations. Pt. 1}
\label{sec:ADMpt1}

\begin{figure}[t]
\centerline{\includegraphics[width=0.8\textwidth]{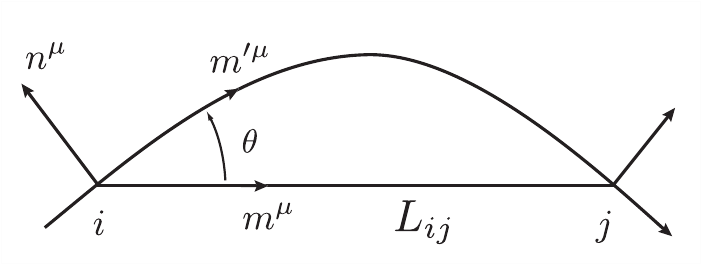}}
\vskip0.5cm
\caption{\normalfont In this figure the lower (straight) curve is the geodesic that joins
$i$ to $j$. The upper curve (which is not shown in figure (\ref{LamEta})) arises from the
intersection of the Cauchy surface with the 2-dimensional surface $S$. The unit vectors $n^\mu$ and
$m'^\mu$ are orthogonal. Note that, in general, $\eta$ is \emph{not} constant on each Cauchy surface.}
\label{Theta}
\end{figure}

\begin{figure}[t]
\centerline{\includegraphics[width=0.8\textwidth]{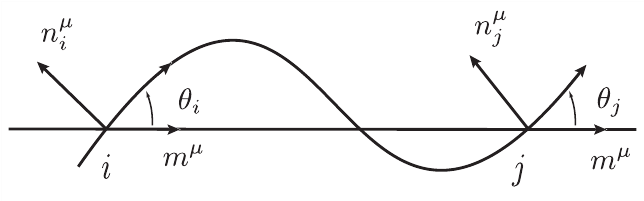}}
\vskip0.5cm
\caption{\normalfont This is a situation that we explicitly exclude. In this case the
points $i$ and $j$ are so far apart that the geodesic intersects the Cauchy surface at points other than
$i$ and $j$. In this case $\theta_i$ and $\theta_j$ have the same signs, contrary to the assumptions
made in the text.}
\label{ThetaPos}
\end{figure}

Consider a typical Cauchy surface $\Sigma$ and suppose that the pair of time like curves $C_i$ and $C_j$
intersect $\Sigma$ at the points $i$ and $j$ respectively. At $i$ we have two vectors $n^\mu$, the unit
normal to $\Sigma$ and $m^\mu$, the unit tangent to the geodesic that connects $i$ to $j$. If we
construct a third unit vector $m'^\mu$ as a linear combination of $n^\mu$ and $m^\mu$,
\[
m'^\mu \cosh\theta = m^\mu + n^\mu \sinh\theta
\]
we can, by careful choice of the boost angle $\theta$, ensure that $m'^\mu$ is tangent to $\Sigma$. That
is, we require $\theta$ such that $0=n_\mu m'^\mu$. This arrangement is shown in figure (\ref{Theta}). In
what follows we will be looking at the behaviour of various expressions in the case where $L_{ij}$ is
small. So our present task is to ask : How does $\theta$ vary with $L_{ij}$? The first observation is
trivial : $\theta\rightarrow0$ as $L_{ij}\rightarrow0$. Now from $0=n_\mu m'^\mu$ we have
\[
\sinh\theta = n_\mu m^\mu
\]
and thus across the leg we have
\[
\left[\sinh\theta\right]_i^j = \left[ n_\mu m^\mu \right]_i^j = \DLijDeta\tag{ErrA}
\label{eqn:ErrorA}
\]
If we choose $L_{ij}$ to be sufficiently small then we can be sure that the geodesic (that joins $i$ to
$j$) intersects $\Sigma$ only at $i$ and $j$ (see figure (\ref{ThetaPos})). From this constraint we
observe that $\theta_i$ and $\theta_j$ must be of opposite signs and thus
\[
  \left\vert\sinh\theta_i\right\vert
+ \left\vert\sinh\theta_j\right\vert
= \left\vert\DLijDeta\right\vert
\]
Thus each term on the left must be of order $\BigO{dL_{ij}/\deta}$, that is
\[
\theta = \BigO{\DLijDeta}\qquad\text{as\ }L_{ij}\rightarrow0
\]

\ErrorDate{01-Jan-2010}

{\bf There is a small error in the above argument.} I should have written
\[
\left[N\sinh\theta\right]_i^j = \left[ \eta_\mu m^\mu \right]_i^j = \DLijDeta
\]
The middle term uses $\eta_\mu$ whereas the previous incorrect expression (\ref{eqn:ErrorA})
uses $n_\mu$. Since $\eta_\mu = N n_\mu$ the correction introduces the lapse function. As a
consquence of this error I now need to take account of the behaviour of the lapse across the
leg. For a short leg the lapse is almost constant and can thus be factored out. This change
should be carried through leading to the final statement that
\[
\theta = \BigO{\frac{1}{N}\DLijDeta}\qquad\text{as\ }L_{ij}\rightarrow0
\]

\ErrorDateEnd

\subsection{The first variation}

Our aim in this section is to recast the expressions for the first and second variations in terms of the
familiar ADM data, the lapse, shift and extrinsic curvatures.

The extrinsic curvature, $K_{\mu\nu}$, can be defined in a number of ways (see \cite{gourg:2007-01}), such
as
\[
N K_{\mu\nu} = - N n_{\mu;\nu} - \bot(N_{,\mu}) n_{\nu}
\]
where $\bot$ is the projection operator
($\bot{}^\mu{}_{\nu} = h^\mu{}_{\nu} = \delta^\mu{}_{\nu} + n^\mu n_\nu$).
Then
\[
\eta^\mu{}_{;\nu} = \left(N n^\mu\right)_{;\nu}
                  = N_{,\nu} n^\mu - \bot(N^{,\mu}) n_\nu - N K^\mu{}_{\nu}
\]
and thus
\begin{spreadlines}{5pt}
\begin{align*}
m^\mu m^\nu\left(N n_{\mu}\right)_{;\nu}
&= m^\mu m^\nu \left(N_{,\nu} n_{\mu} + N n_{\mu;\nu}\right)\\
&= \left(m^\nu N_{,\nu}\right)\sinh\theta - NK_{\mu\nu}m^\mu m^\nu\label{eqn:ErrorB}\tag{ErrB}
\end{align*}
\end{spreadlines}

\ErrorDate{01-Jan-2010}

{\bf There is another small error here.} The last line two lines should be
\begin{spreadlines}{5pt}
\begin{align*}
m^\mu m^\nu\left(N n_{\mu}\right)_{;\nu}
&= m^\mu m^\nu \left(N_{,\nu} n_{\mu} + N n_{\mu;\nu}\right)\\
&= -\left(n^\nu N_{,\nu}\right)\sinh^2\theta - NK_{\mu\nu}m^\mu m^\nu
\end{align*}
\end{spreadlines}
This error is carried through into the next two equations (\ref{eqn:ErrorC}) and
(\ref{eqn:ErrorD}) but the remaining equations in this section are correct. This also
means that the error term in (\ref{ADMDLija}) should be $\BigO{L^3}$ which works in our
favour.

\ErrorDateEnd

This can now be substituted into the integral for the first variation \eqref{FirstDeriv}
\begin{spreadlines}{5pt}
\begin{align*}
\DLijDeta
&= \int_i^j\> m^\mu m^\nu \eta_{\mu;\nu} ds\\
&= \int_i^j\> \left( m^\nu N_{,\nu}\sinh\theta - N K_{\mu\nu} m^\mu m^\nu \right)\>ds
\label{eqn:ErrorC}\tag{ErrC}
\end{align*}
\end{spreadlines}
Recall that we are dealing with short geodesic segments. Thus we can use any of a number of methods to
estimate the integral. To be specific, we will chose a mid point rule (see \cite{num-recipes:2007}) which
leads to
\[
\DLijDeta = \left( m^\nu N_{,\nu}\sinh\theta\right) L_{ij}
          - \left( N K_{\mu\nu} m^\mu m^\nu\right) L_{ij}
+ \BigO{L^2}
\label{eqn:ErrorD}\tag{ErrD}
\]
where each term is evaluated at the mid-point of the geodesic.
But since $\theta =\BigO{dL/\deta}$ we see that the first term is of order $\BigO{L^2}$
and thus
\begin{equation}
\DLijDeta = - \left( N K_{\mu\nu} m^\mu m^\nu\right) L_{ij} + \BigO{L^2}
\label{ADMDLija}
\end{equation}
Notice that $m^\mu$ is the unit tangent vector at the mid-point of the geodesic that joins $i$ to $j$ and
thus we have
\begin{equation}
m^\mu L_{ij} = x^\mu_j - x^\mu_i + \BigO{L^3}
\label{mLijDx}
\end{equation}
So, if we put $\Delta x^\mu_{ij} = x^\mu_j - x^\mu_i$ we can rewrite \eqref{ADMDLija} as
\begin{equation}
\DLsqijDeta = -2 N K_{\mu\nu} \Delta x^\mu_{ij} \Delta x^\nu_{ij} + \BigO{L^3}
\label{ADMDLijb}
\end{equation}

\subsection{The second variation}

Once again we use the basic definition of the extrinsic curvature to express the terms appearing in the
second variation in an ADM form. We will do the calculations by splitting our previous expression for
the second variation \eqref{SecondDeriv} into the following pieces
\begin{spreadlines}{5pt}
\begin{gather*}
\DDLijDeta = J_1 + J_2 + J_3 + J_4\\
J_1 = \left[ \eta^\alpha{}_{;\mu} \eta^\mu m_\alpha \right]_i^j \qquad
J_2 = - \int_i^j\> R_{\mu\alpha\nu\beta} m^\mu m^\nu \eta^\alpha \eta^\beta \>ds\\
J_3 =\int_i^j\> \eta_{\mu;\alpha}\eta^{\mu}{}_{;\nu}m^\alpha m^\nu \>ds\qquad
J_4 = - \int_i^j\> \left(m_\mu m^\nu \eta^\mu{}_{;\nu}\right)^2 \>ds
\end{gather*}
\end{spreadlines}

\subsubsection{The second term}

We start with this term as it requires very little work. We simply substitute $\eta^\mu = Nn^\mu$ and
approximate the integral via a mid-point rule leading to
\[
J_2 = - \int_i^j\> R_{\mu\alpha\nu\beta} m^\mu m^\nu \eta^\alpha \eta^\beta \>ds
= - N^2 R_{\mu\alpha\nu\beta} m^\mu m^\nu n^\alpha n^\beta L_{ij} +\BigO{L^2}
\]

\subsubsection{The fourth term}

Here we use $m^\mu m^\nu \eta_{\mu;\nu} = -N K_{\mu\nu}m^\mu m^\nu + \BigO{L}$ (the error term
arises from the $n_\mu m^\mu=\sinh\theta=\BigO{L}$ terms). Thus we are led to
\begin{spreadlines}{5pt}
\begin{align*}
J_4 = - \int_i^j\> \left( m^\mu m^\nu \eta_{\mu;\nu} \right)^2 \>ds
 &= - \int_i^j\> \left( -N K_{\mu\nu}m^\mu m^\nu + \BigO{L} \right)^2 \>ds\\
 &= - \left(N K_{\mu\nu}m^\mu m^\nu\right)^2 L_{ij} + \BigO{L^2}\\
 &= - \frac{1}{\Lij} \left( \DLijDeta \right)^2  + \BigO{L^2}
\end{align*}
\end{spreadlines}
where we have used \eqref{ADMDLija} in the second last line.

The remaining terms are not so easily dealt with.

\subsubsection{The third term}

For the third term the details are as follows
\begin{spreadlines}{5pt}
\begin{align*}
J_3 &= \int_i^j\> \eta_{\mu;\alpha}\eta^{\mu}{}_{;\nu}m^\alpha m^\nu ds\\
&= \int_i^j\> \left( N_{,\alpha} n_\mu + N n_{\mu;\alpha} \right)
              \left( N_{,\beta} n^\mu + N n^\mu{}_{;\beta}\right) m^\alpha m^\beta\>ds\\
&= \int_i^j\> \left( -\left(N_{,\alpha} m^\alpha\right)^2
              + N^2 n_{\mu;\alpha} n^\mu{}_{;\beta} m^\alpha m^\beta \right)\>ds\\
&= \int_i^j\> \left( -\left(N_{,\alpha} m^\alpha\right)^2
              +\left(\bot(N_{,\mu})n_\alpha + K_{\mu\alpha}\right)
               \left(\bot(N^{,\mu})n_\beta+K^\mu{}_{\beta}\right) m^\alpha m^\beta \right)\>ds\\
&=  \int_i^j\> \left( -\left(N_{,\alpha} m^\alpha\right)^2
                      + N^2 K_{\mu\alpha} K^\mu{}_{\beta} m^\alpha m^\beta + \BigO{L} \right) \>ds
\end{align*}
\end{spreadlines}
\ErrorDate{12 Mar 2011}
The second last line in the above equation should read
\begin{equation*}
\phantom{J_3} = \int_i^j\> \left( -\left(N_{,\alpha} m^\alpha\right)^2
              +\left(\bot(N_{,\mu})n_\alpha + NK_{\mu\alpha}\right)
               \left(\bot(N^{,\mu})n_\beta+NK^\mu{}_{\beta}\right) m^\alpha m^\beta \right)\>ds
\end{equation*}
Notice the two extra $N$'s. The final line in the above equation is correct.
\ErrorDateEnd
The error term $\BigO{L}$ in the last line arises from terms of the form $n_\mu m^\mu=\sinh\theta=\BigO{L}$.
Now we use the mid-point rule, once again, to obtain
\[
J_3 = \int_i^j\> \eta_{\mu;\alpha}\eta^{\mu}{}_{;\nu}m^\alpha m^\nu ds
= -\left(N_{,\alpha} m^\alpha\right)^2 L_{ij}
  + N^2 K_{\mu\alpha} K^\mu{}_{\beta} m^\alpha m^\beta L_{ij} + \BigO{L^2}
\]

\subsubsection{The first term}

Finally, we turn to the first term $\left[ \eta^\alpha{}_{;\mu} \eta^\mu m_\alpha \right]_i^j$.
Using the same substitutions as we have used before and also using $Nn^\mu N_{,\mu} = dN/\deta$ we
obtain
\[
J_1 = \left[ \eta^\alpha{}_{;\mu} \eta^\mu m_\alpha \right]_i^j
= \left[ \frac{1}{N}\frac{dN}{\deta} \eta^\mu m_\mu \right]_i^j
+ \left[ N N_{,\mu} m^\mu \right]_i^j
\]
We choose to write this result as a sum of two terms each of the form $[\cdots]_i^j$ so that we
can deal with each term separately. In the first term we have $(1/N)dN/\deta$ which varies slowly
over the short geodesic and thus may be taken as a constant (plus an error term of order $\BigO{L}$),
thus we have
\begin{spreadlines}{5pt}
\begin{align*}
\left[ \frac{1}{N}\frac{dN}{\deta} \eta^\mu m_\mu \right]_i^j
&= \frac{1}{N}\frac{dN}{\deta} \left[ \eta^\mu m_\mu \right]_i^j
+ \left[ \eta^\mu m_\mu \right]_i^j \BigO{L}\\
&=\frac{1}{N}\frac{dN}{\deta} \DLijDeta + \BigO{L^2}
\end{align*}
\end{spreadlines}
For the second term we use a Taylor series expansion
\begin{spreadlines}{5pt}
\begin{align*}
\left[ N N_{,\mu} m^\mu \right]_i^j
&= \frac{d\ }{ds}\left(N N_{,\mu} m^\mu\right) L_{ij} + \BigO{L^2}\\
&= \left(N N_{,\mu} m^\mu\right)_{;\alpha} m^\alpha L_{ij} + \BigO{L^2}\\
&= \left(N_{,\mu} m^\mu\right)^2 L_{ij} + N N_{;\alpha\beta} m^\alpha m^\beta L_{ij} + \BigO{L^2}
\end{align*}
\end{spreadlines}
The appearance of the term $N_{;\alpha\beta}$ is encouraging -- it reminds us of the similar term in the
ADM equations. We can improve on this situation. Notice that $m'^\mu = m^\mu + \BigO{L}$ and thus
\[
N_{;\alpha\beta} m^\alpha m^\beta = N_{;\alpha\beta} m'^\alpha m'^\beta + \BigO{L}
\]
However, $m'^\mu$ is tangent to $\Sigma$ thus we also have
\[
N_{;\alpha\beta} m^\alpha m^\beta
= N_{|\alpha\beta} m'^\alpha m'^\beta + \BigO{L}
= N_{|\alpha\beta} m^\alpha m^\beta + \BigO{L}
\]
where the vertical stroke denotes covariant differentiation with respect to the 3-metric intrinsic to
$\Sigma$.

Combining these two results we obtain our final estimate for the first term in the second variation
\[
\left[ \eta^\alpha{}_{;\mu} \eta^\mu m_\alpha \right]_i^j
= \frac{1}{N}\frac{dN}{\deta} \DLijDeta
+ \left(N_{,\mu} m^\mu\right)^2 L_{ij} + N N_{|\alpha\beta} m^\alpha m^\beta L_{ij}
+ \BigO{L^2}
\]

Now we can reassemble the pieces. The result is
\begin{spreadlines}{8pt}
\begin{align*}
\DDLijDeta &= \frac{1}{N}\frac{dN}{\deta} \DLijDeta
            - \frac{1}{\Lij} \left( \DLijDeta \right)^2
            + N^2 K_{\mu\alpha}K^\mu{}_{\beta} m^\alpha m^\beta L_{ij}\\
          & + N N_{|\alpha\beta} m^\alpha m^\beta L_{ij}
            - N^2 R_{\mu\alpha\nu\beta} m^\mu m^\nu n^\alpha n^\beta L_{ij}
            + \BigO{L^2}
\end{align*}
\end{spreadlines}
We are almost finished, we just need to do a little bit of tidying up. We multiply both sides by
$L_{ij}/N$ and noting that
\def\Dxij{\Delta x_{ij}}
\def\ADMDDLsqDeta{\frac{d\ }{\deta}\left(\frac{1}{N}\DLsqijDeta\right)}
\begin{spreadlines}{8pt}
\begin{gather*}
\DDLsqijDeta = 2\left(\DLijDeta\right)^2 + 2\Lij \DDLijDeta\\
\ADMDDLsqDeta =
-\frac{1}{N^2} \frac{dN}{\deta} \DLsqijDeta
+\frac{1}{N} \DDLsqijDeta
\end{gather*}
\end{spreadlines}
we can rewrite the above equation as
\begin{align}
\label{ADMDLijc}
\ADMDDLsqDeta &= 2 N_{|\alpha\beta} \Dxij^\alpha \Dxij^\beta\\
              & \quad+ 2 N\left( K_{\mu\alpha}K^\mu{}_\beta - R_{\mu\alpha\nu\beta} n^\mu n^\nu \right)
                      \Dxij^\alpha \Dxij^\beta + \BigO{L^3}\notag
\end{align}
where we have also used $\Delta x^\mu_{ij} = m^\mu L_{ij} + \BigO{L^3}$.

For completeness, we repeat here the result we previously obtained for the first time derivative,
\begin{equation}
\DLsqijDeta = -2 N K_{\mu\nu} \Delta x^\mu_{ij} \Delta x^\nu_{ij} + \BigO{L^3}
\tag{\ref{ADMDLijb}}
\end{equation}

\section{The ADM evolution equations. Pt. 2}
\label{sec:ADMpt2}

This completes the first stage of the construction. We have successfully expressed the first and second
variations in terms of the extrinsic and Riemann curvatures. Our second and final stage will, among
other things, introduce the metric tensor as a replacement for the geodesic arc-lengths. As we shall soon
see, this is not a difficult task. The most notable change is not in the symbols, from $L^2_{ij}$ to
$g_{\mu\nu}$, but in the structure of the equations. We will be re-working an equation defined
over a geodesic segment into an new equation defined at a point.

Consider a typical geodesic segment with end-points $i$ and $j$. The time like worldlines $C_i$ and $C_j$
generated by $i$ and $j$ are, by assumption, orthogonal to the Cauchy surfaces. Thus we can use these
curves to propagate the spatial coordinates of each Cauchy surface forward in time. This means that
the spatial coordinates of any point on $C_i$ are constant along $C_i$ and thus $0=d\Delta
x^\mu_{ij}/\deta$.

We now introduce the metric by estimating $L_{ij}$ using a mid-point rule for $\int ds$,
\begin{spreadlines}{5pt}
\begin{align*}
L_{ij} &= \int_i^j\>\left( g_{\mu\nu} \PxPlam\mu \PxPlam\nu \right)^{1/2} M\>\dlam\\
       &= \left( g_{\mu\nu} \PxPlam\mu \PxPlam\nu \right)^{1/2} M + \BigO{L^2}
\end{align*}
\end{spreadlines}
where each term on the right hand side is evaluated at the mid point of the geodesic. But we have
previously shown
(\ref{vecNMa},\ref{vecNMb}) and \eqref{mLijDx} that $\partial x^\mu/\partial\lambda = m^\mu L_{ij} =
\Delta x^\mu_{ij} +\BigO{L^3}$. We can use this to estimate $L^2_{ij}$ as
\[
L^2_{ij} = g_{\mu\nu} \Delta x^\mu_{ij} \Delta x^\nu_{ij} + \BigO{L^3}
\]
We can go one step further by noting that $g_{\mu\nu} = h_{\mu\nu} - n_\mu n_\nu$ and
$n_\mu\Delta x^\mu_{ij} = L_{ij}\sinh\theta = \BigO{L^2}$ and thus to leading order in $L$ we have
\[
  g_{\mu\nu}\Delta x^\mu_{ij} \Delta x^\nu_{ij}
= h_{\mu\nu}\Delta x^\mu_{ij} \Delta x^\nu_{ij} + \BigO{L^4}
\]
which, when substituted into the above, leads to
\begin{equation}
L^2_{ij} = h_{\mu\nu} \Delta x^\mu_{ij} \Delta x^\nu_{ij} + \BigO{L^3}
\label{hmnLSQij}
\end{equation}
It is now just a short step to the finish line. First substitute \eqref{ADMDLijb} into \eqref{ADMDLijc}
and then \eqref{hmnLSQij} into (\ref{ADMDLijb}) and finally take the $\Delta x^\mu_{ij}$ terms out
through the time derivatives. Then notice that the $\Delta x^\mu_{ij}$ are arbitrary and that the
coefficients of $\Delta x^\mu_{ij}\Delta x^\nu_{ij}$ are symmetric in $\mu\nu$ and purely spatial. This
allows us to cancel the $\Delta x^\mu$ from both sides of the equations after which we can safely let
$L\rightarrow0$ (the details of this series of substitutions and eliminations are excluded as they
follow very standard lines). As expected the final result is nothing other than the familiar ADM
evolution equations with a zero shift vector
\begin{spreadlines}{8pt}
\begin{align*}
\frac{dh_{\mu\nu}}{\deta} &= -2 N K_{\mu\nu}\\
\frac{dK_{\mu\nu}}{\deta} &= -N_{|\mu\nu}
              - N\left( K_{\mu\alpha}K^\alpha{}_\nu - R_{\mu\alpha\nu\beta} n^\alpha n^\beta \right)
\end{align*}
\end{spreadlines}

\appendix

\section{The first variation}
\label{app:1stvar}

We know that the mixed partial derivatives of $x^\mu(\lambda,\eta)$ must commute, thus we must have
\begin{gather*}
  \lambda^\mu{}_{,\nu}\eta^\nu = \eta^\mu{}_{,\nu}\lambda^\nu
\end{gather*}
and for a symmetric connection (which we are using) we also have
\begin{gather*}
  \lambda^\mu{}_{;\nu}\eta^\nu = \eta^\mu{}_{;\nu}\lambda^\nu
\end{gather*}
which can be re-expressed, terms of the unit vectors $n^\mu$ and $m^\mu$, as
\begin{gather}
  \left(N n^\mu\right)_{;\nu}\left(M m^\nu\right)
= \left(M m^\mu\right)_{;\nu}\left(N n^\nu\right)
\label{NMcommute}
\end{gather}
Finally, as the vector $m^\mu$ is the unit tangent to an $\eta=\>$constant geodesic, we have
\[
0 = m^\mu{}_{;\nu} m^\nu
\]
and
\[
0 = \frac{\partial^2 x^\mu}{\partial\lambda^2}
  + \Gamma^\mu_{\alpha\beta} \PxPlam{\alpha} \PxPlam{\beta}
\]

We will use the above equations frequently in the following discussions.

Here we consider the geodesic arc-length and its first time derivative,
\begin{spreadlines}{5pt}
\begin{align*}
\Lij &= \int_0^1\>\frac{ds}{\dlam}\>\dlam
     = \int_0^1\>\left( g_{\mu\nu} \PxPlam{\mu}\PxPlam{\nu} \right)^{1/2}\>\dlam\\[5pt]
\DLijDeta &= \frac{d\>}{\deta} \int_0^1\>\frac{ds}{\dlam}\>\dlam
           =\int_0^1\>\frac{\partial\>}{\partial\eta}
               \left( g_{\mu\nu} \PxPlam{\mu}\PxPlam{\nu} \right)^{1/2}\>\dlam
\end{align*}
\end{spreadlines}
Note that the path $x^\mu(\lambda,\eta)$ in each of these integrals is a geodesic and that
$\eta$ is constant along the geodesic. The second integral in the last equation above can be
readily evaluated using standard techniques (expand the $\eta$ derivative, swap orders of
mixed derivatives, integrate by parts and impose the geodesic equation). The result is
\[
\DLijDeta = \frac{1}{\Lij} \left[ g_{\mu\nu}\PxPlam{\mu}\PxPeta{\nu} \right]_i^j
          = \left[ m_\mu \eta^\mu \right]_i^j
\]
where we have taken the small liberty of replacing the limits 0 and 1 with the more suggestive
labels $i$ and $j$. This is an elegant result -- it shows that for a geodesic segment,
$d\Lij/\deta$ can be computed from data defined only at the end points of the geodesic. This
may seem simple but it hides a significant complexity -- the data involved can only be found
by solving a two-point boundary value problem.

Despite this compact and elegant form for the first time derivative, we will now develop an
alternative integral expression that happens to be better suited to our later calculations of the
second time derivative. Consider for the moment the quantity $Q$ defined by
\[
Q = \int_i^j\>m_\mu m^\nu \left(N n^\mu\right)_{;\nu}\>ds
\]
with the integration path being, as expected, an $\eta=\>$constant geodesic. We will now show
that $Q = d\Lij/\deta$. We begin by writing $ds=(ds/\dlam)\dlam=M\dlam$ and using the
commutation relation \eqref{NMcommute} to obtain
\[
Q = \int_i^j\>m_\mu \eta^\nu \left(Mm^\mu\right)_{;\nu}\>\dlam
\]
Now expand the covariant derivative and use $1=m_\mu m^\mu$ and $0=m_\mu m^\mu{}_{;\nu}$ to
obtain
\begin{align*}
Q = \int_i^j\> \eta^\nu M_{,\nu}\>\dlam
  = \int_i^j\> \frac{\partial\>}{\partial\eta}\left(\frac{ds}{\dlam}\right)\>\dlam
  = \frac{d\>}{\deta}\int_i^j\>\frac{ds}{\dlam}\>\dlam
  = \DLijDeta
\end{align*}
Thus we have shown that
\begin{equation}
\DLijDeta 
          = \left[ m_\mu \eta^\mu \right]_i^j
          = \int_i^j\>m_\mu m^\nu \eta^\mu{}_{;\nu}\>ds
\tag{\ref{FirstDeriv}}
\end{equation}

Our challenge now is to compute the second time derivative. This proceeds in a manner similar
to the above calculation though it is a tad lengthy.

\section{The second variation}
\label{app:2ndvar}

\def\subsection#1{\vskip 12pt plus 4pt minus 2pt\leftline{\bf #1}\vskip0pt plus 2pt minus 1pt}

To compute the second derivative we need only apply $d/\deta$ to \eqref{FirstDeriv}. This
leads to
\begin{spreadlines}{5pt}
\begin{align*}
\DDLijDeta &= \frac{d\>}{\deta}\int_i^j\>m_\mu m^\nu \eta^\mu{}_{;\nu}\>ds
            = \int_i^j\>\frac{\partial\>}{\partial\eta}
             \left( m_\mu m^\nu \eta^\mu{}_{;\nu} M \right)\>\dlam\\[5pt]
&= \int_i^j\>\left( m_\mu \eta^\mu{}_{;\nu} \lambda^\nu \right)_{;\alpha} \eta^\alpha\>\dlam\\[2pt]
&= \int_i^j\>\left(  m_{\mu;\alpha}\eta^\mu{}_{;\nu}\lambda^\nu
            + m_\mu\eta^\mu{}_{;\nu;\alpha} \lambda^\nu
            + m_\mu \eta^\mu{}_{;\nu} \lambda^\nu{}_{;\alpha} \right) \eta^\alpha\>\dlam
\end{align*}
\end{spreadlines}
We will apply various manipulations to the three main parts of this integral and we will make
extensive use of the geodesic equations, $0=m^\mu{}_{;\nu} m^\nu$, the commutation relations,
$\lambda^\mu{}_{;\nu}\eta^\nu = \eta^\mu{}_{;\nu}\lambda^\nu$ and the observations that
$m^\mu$ is a unit vector along the geodesic.

We start by splitting the above integral into three pieces
\begin{spreadlines}{5pt}
\begin{align*}
I_1 & = \int_i^j\> m_{\mu;\alpha}\eta^\mu{}_{;\nu}\lambda^\nu \eta^\alpha\>\dlam\\
I_2 & = \int_i^j\> m_\mu\eta^\mu{}_{;\nu;\alpha} \lambda^\nu \eta^\alpha\>\dlam\\
I_3 & = \int_i^j\> m_\mu \eta^\mu{}_{;\nu} \lambda^\nu{}_{;\alpha} \eta^\alpha\>\dlam
\end{align*}
\end{spreadlines}
which we will now attempt to simplify.

\subsection{Integral $I_1$}

Put $\lambda^\nu = m^\nu M$ and $m_{\mu;\alpha} M = (m_\mu M)_{;\alpha} - m_\mu M_{;\alpha}$
and then use the commutation rule on $\lambda_{\mu;\alpha}\eta^\alpha$ to
obtain
\[
I_1 = \int_i^j\> \eta_{\mu;\alpha}\eta^{\mu}{}_{;\nu}m^\alpha m^\nu M\>\dlam
    - \int_i^j\> m_\mu M_{;\alpha}\eta^\alpha \eta^\mu{}_{;\nu}m^\nu\>\dlam
\]
Consider the second integral in this pair and denote it by $I_4$. Since $m^\mu$ is a unit
vector we can slide a factor of $m_\theta m^\theta$ inside $M_{;\alpha}$, like
this
\begin{align*}
I_4 &= \int_i^j\> m_\mu \left(m_\theta m^\theta M\right)_{;\alpha}
         \eta^\alpha \eta^\mu{}_{;\nu}m^\nu\>\dlam\\[5pt]
   &= \int_i^j\> m_\mu \left(  m_{\theta;\alpha} m^\theta M
                             + m_{\theta}\left(m^\theta M\right)_{;\alpha} \right)
                       \eta^\alpha \eta^\mu{}_{;\nu} m^\nu\>\dlam
\end{align*}
The term $m_{\theta;\alpha} m^\theta$ is zero since $m^\mu$ is a unit vector while the
remaining term is ripe for a commutation operation. This leads to
\begin{align*}
I_4 = \int_i^j\> m_\mu m_\theta \eta^\theta{}_{;\alpha}
                  m^\alpha M \eta^\mu{}_{;\nu} m^\nu\>\dlam
    = \int_i^j\> \left(m_\mu m^\nu \eta^\mu{}_{;\nu}\right)^2 M \>\dlam
\end{align*}
So our final expression for $I_1$ is
\[
I_1 = \int_i^j\> \left( \eta_{\mu;\alpha}\eta^{\mu}{}_{;\nu}m^\alpha m^\nu
                       - \left(m_\mu m^\nu \eta^\mu{}_{;\nu}\right)^2 \right) M \>\dlam
\]

\subsection{Integral $I_3$}

We step out of sequence here because one term arises in this computation that will be useful
when we tackle the second integral $I_2$.

This integral is slightly easier to work with than the first integral and it will give rise to
the Riemann tensor. The main device used here is to swap the order of the second partial
derivatives on $\eta^\mu{}_{;\nu;\alpha}$ balanced by the addition of the Riemann tensor.
Thus we have
\begin{spreadlines}{5pt}
\begin{align*}
I_3 &= \int_i^j\> m_\mu \eta^\mu{}_{;\nu;\alpha} m^\nu \eta^\alpha M \>\dlam\\
    &= \int_i^j\> m_\mu \left( \eta^\mu{}_{;\alpha;\nu}
                               + R^\mu{}_{\rho\alpha\nu}\eta^\rho
                         \right) m^\nu \eta^\alpha M \>\dlam\\
    &= I_5 - \int_i^j\> R_{\mu\alpha\nu\beta} m^\mu m^\nu \eta^\alpha \eta^\beta M \>\dlam
\end{align*}
\end{spreadlines}
where we have introduced a fifth integral,
\[
I_5 = \int_i^j\> m_\mu \eta^\mu{}_{;\alpha;\nu}m^\nu \eta^\alpha M \>\dlam
\]

\subsection{Integral $I_2+I_5$}

As we shall soon see, the integrand for $I_2+I_5$ can be combined to form a total derivative and thus the
integration is trivial. We start by forming the sum $I_2$ and $I_5$
\[
I_2 + I_5 = \int_i^j\>\left( \eta_{\alpha;\mu} \eta^\mu{}_{;\beta} m^\alpha m^\beta
                            + \eta^\mu{}_{;\alpha;\nu} m_\mu \eta^\alpha m^\nu
                       \right)\> ds
\]
where $ds = M \dlam$. By careful inspection of the integrand, while noting the geodesic conditions,
$0=m^\mu{}_{;\nu}m^\nu$, it is not hard to see that the integrand can also be written as
$\left(\eta^\alpha{}_{;\mu}\eta^\mu m_\alpha\right)_{;\nu} m^\nu$.
Thus we have
\begin{spreadlines}{5pt}
\begin{align*}
I_2+I_5 &= \int_i^j\>\left(\eta^\alpha{}_{;\mu}\eta^\mu m_\alpha\right)_{;\nu} m^\nu\>ds\\
        &= \left[ \eta^\alpha{}_{;\mu} \eta^\mu m_\alpha \right]_i^j
\end{align*}
\end{spreadlines}

Our job is done, all of the integrals have been evaluated as far as possible -- all that
remains is to combine the above results. This leads to
\begin{spreadlines}{5pt}
\begin{align}
\DDLijDeta
   &= \left[ \eta^\alpha{}_{;\mu} \eta^\mu m_\alpha \right]_i^j
    - \int_i^j\> R_{\mu\alpha\nu\beta} m^\mu m^\nu \eta^\alpha \eta^\beta \>ds\notag\\
   &+\int_i^j\> \left(
          \eta_{\mu;\alpha}\eta^{\mu}{}_{;\nu}m^\alpha m^\nu
          - \left(m_\mu m^\nu \eta^\mu{}_{;\nu}\right)^2 \right) \>ds
\tag{\ref{SecondDeriv}}
\end{align}
\end{spreadlines}

This last integral can be simplified slightly by introducing
\[
v^\mu = \eta^\mu - \eta_\rho m^\rho m^\mu
\]
which leads to
\begin{spreadlines}{5pt}
\begin{equation*}
\DDLijDeta
    = \left[ \eta^\alpha{}_{;\mu} \eta^\mu m_\alpha \right]_i^j
    - \int_i^j\> R_{\mu\alpha\nu\beta} m^\mu m^\nu \eta^\alpha \eta^\beta \>ds
    + \int_i^j\> v_{\mu;\alpha}v^{\mu}{}_{;\nu}m^\alpha m^\nu \>ds
\end{equation*}
\end{spreadlines}

\clearpage

\captionsetup{margin=0pt,font=small,labelfont=bf}

\clearpage


\providecommand{\href}[2]{#2}\begingroup\raggedright\endgroup

\end{document}